# TURKISH TEXT RETRIEVAL EXPERIMENTS USING LEMUR TOOLKIT

Kutlu Emre Yılmaz, Ahmet Arslan, Ozgur Yilmazel[1]

[1]*Anadolu University, Computer Engineering Department*
*Eskisehir, Turkey*

**ABSTRACT**

We used Lemur Toolkit, an open source toolkit designed for Information Retrieval (IR) research, for our automated indexing and retrieval experiments on a TREC-like test collection for Turkish. We study and compare three retrieval models Lemur supports, especially Language modeling approach to IR, combined with language specific preprocessing techniques. Our experiments show that all retrieval models benefits from language specific preprocessing in terms of retrieval quality. Also Language Modeling approach is the best performing retrieval model when language specific preprocessing applied.

**KEYWORDS**

Turkish Information Retrieval; Lemur Toolkit; Language Modeling

## 1. INTRODUCTION

Information retrieval is a broad science with expanding subfields, while most new approaches to IR field are first introduced using English document collections, IR experiments in other languages also move parallel to current trends. There are some studies on Turkish IR research incorporating widely used vector space and probabilistic retrieval models combined with different language specific preprocessing. However, there is no published study comparing effectiveness of language modeling approach for Turkish text retrieval.

In this paper we compare retrieval performance of Turkish as an agglutinative language with productive inflectional and derivational suffixations, both from language modeling approach and conventional retrieval approaches. For this purpose three main retrieval models were used in our experiments. They are Lemur TF-IDF, OKAPI and Language Modeling. Also we investigate how language specific properties of Turkish affect retrieval performance.

Our experiments are conducted using open source Lemur Toolkit and based on a TREC-like Turkish test collection consisting of 485,000 documents, 72 ad-hoc queries and relevance judgments. We made experiments with different settings to find optimum parameters and also tried to do language specific improvements for all three retrieval models.

## 2. RELATED WORK

Different researches have conducted on large text collections and different retrieval models were proposed. Vector space [7] model represents documents and queries as high dimensional vectors. K. S. Jones [4] showed that using inverse document frequency and term frequency together as a term weighing method is much better than using term frequency alone. Also, OKAPI system [8, Section 2 and 3], a probabilistic model, evolved at the TREC conferences, integrates document length normalization factor into term weighing methods. In addition to these conventional models, J. Ponte and W. B. Croft [5] proposed a model which scores documents according to the probability of query generated by document language model. C. Zhai and J. Lafferty [9] examined effects of different smoothing parameters on language modeling.

## 3. EXPERIMENTAL SETUP

In this study Milliyet Collection [3] created by F. Can and Bilkent Information Retrieval Group is used. Milliyet Collection consists of 408,305 news articles, 72 ad-hoc queries and corresponding relevance judgments.

All experiments are conducted using Lemur[1] Toolkit which is an open source IR research system developed collaboratively by University of Massachusetts, Amherst and Carnegie Mellon University. Experiments on this paper are based on three retrieval methods. Main retrieval model is a unigram language-modeling algorithm which ranks documents by similarity of document and query language models using Kullback-Leibler [2] divergence as a measure. Three interpolation based smoothing methods [9] are available at Lemur[1] are used for language modeling algorithm. Other two retrieval models are OKAPI retrieval algorithm [6] [8] and a dot product function [10] using a TF-IDF variant for term weighing.

In this work no stemming plus two stemming algorithms are used. An affix stemmer and Zemberek stemmer. Implementation of affix stemmer is done by Evren (Kapusuz) Çilden in Snowball[2] language and is freely available. Zemberek[3] is an open source Natural Language Processing (NLP) library designed for Turkic languages especially for Turkish. It provides root forms of given words using a root dictionary-based parser combined with NLP algorithms.

## 4. EXPERIMENTAL RESULTS

For evaluation of ranked retrieval results we used *bpref* [1] metric which is introduced by C. Buckley and E. M. Voorhees in SIGIR 2004. A comparison of three retrieval models based on best *bpref* values is given in Table 1.

Table 1. Best *bpref* values of three retrieval methods with three stemming options.

|  |  | No Stemming | Affix Stemmer | Zemberek Stemmer |
|---|---|---|---|---|
| Lemur TF-IDF | | **0.4324** | **0.5130** | 0.5096 |
| OKAPI | | 0.4230 | 0.5068 | 0.5138 |
| Language Modeling | Jelinek-Mercer | 0.3933 ( = 0.3) | 0.4808 ( = 0.5) | 0.4842 ( = 0.4) |
| | Dirichlet | 0.4206 ( = 2000) | 0.5063 ( = 1000) | **0.5148** ( = 500) |
| | Absolute Discounting | 0.4007 ( = 0.75) | 0.4869 ( = 0.7) | 0.4916 ( = 0.7) |

Our evaluations using Lemur TF-IDF model show that in no stemming applied tests we get best *bpref* values using parameters $k_1=1$, $k_3=1000$, $b=0.2$. When Affix and Zemberek stemmers applied we get best *bpref* values using parameters $k_1=1$, $k_3=1000$, $b=0.4$. Test runs with stemming have higher $b$ values than without stemming runs. That means more document length normalization needed, due to the fact that stemming operation decreased count of unique words in the collection and led to higher term frequencies as it is in long documents.

When we study with OKAPI model we get best results with Zemberek stemmer. In no stemming applied tests we get best *bpref* values using parameters $k_1=1.4$, $k_3=1000$, $b=0.1$. Runs with Affix and Zemberek stemmers applied we get best *bpref* values using parameters $k_1=1$, $k_3=1000$, $b=0.75$. Same as in Lemur TF-IDF model when stemming applied increasing document length normalization constant $b$ gives better results.

In language modeling approach we compared *bpref* values of three smoothing methods with three stemming options using interpolation based smoothing strategy. Bayesian smoothing using Dirichlet priors is the best performing method, absolute discounting is the second and Jelinek-Mercer is the worst performing in our test runs with optimum parameter values. When the best results are taken as measure Bayesian smoothing has best *bpref* values around prior value of 2000 without stemming and around prior value of 1000 with stemming. Absolute discounting has best *bpref* values around delta value of 0.8 without stemming and around 0.7 with stemming.

---

[1] http://www.lemurproject.org/lemur/retrieval.php
[2] http://snowball.tartarus.org/algorithms/turkish/stemmer.html
[3] http://code.google.com/p/zemberek/

Conversely, Jelinek-Mercer has best bpref values around delta value of 0.3 without stemming and around 0.5 with stemming. The relation between optimum smoothing parameters and stemming based on best *bpref* values is given in last three rows of Table 1. We see that language modeling using Bayesian smoothing is the best performing among three smoothing methods.

As C. Zhai and J. Lafferty [9] explained in their paper, smoothing of the document language model shows some similarities with traditional heuristics, such as TF-IDF weighing and document length normalization. The same similarity they mentioned is seen in our studies too. In addition to their implications, the effect of stemming in Turkish IR is similar in two conventional retrieval models and language modeling approach.

## 5. CONCLUSION

IR models we used in our experiments meet at the same points directly or indirectly, the importance of term frequency, inverse document frequency, stemming and document length normalization. We investigated effects of these key concepts in our experiments, especially for language modeling on Turkish IR.

All three IR models have similar responses to the key points mentioned above. Stemming applied experiments in all models give up to %20 performance improvements. Our results are clear evidence of the importance of stemming on Turkish IR and are also a clue for other agglutinative languages. A lemmatizer based stemmer (Zemberek) which uses morphological rules of Turkish gives best results. Language modeling is the best performing retrieval model with Zemberek stemmer, based on *bpref* values. When no special linguistic preprocessing applied Lemur TF-IDF is the best performing retrieval model.

Behavior of language modeling is also similar to other two models (Lemur TF-IDF, OKAPI). In addition to this, as shown in Table 1, while Dirichlet is the third best performing by *bpref* values in no stemming run, it is the first (best performing) with Zemberek stemmer. This might be a sign of language specific dependency of language modeling framework. However, the lack of different test collections on Turkish IR prevents us from having certain conclusions since it is based on document and collection specific characteristics.